\documentclass[runningheads]{llncs}
\usepackage[T1]{fontenc}
\usepackage{graphicx}

\usepackage{booktabs} 
\usepackage{algorithm}
\usepackage{algpseudocode}
\usepackage{bm}
\usepackage{siunitx}
\usepackage{tikz}
\usepackage{adjustbox}
\usepackage[edges]{forest}
\usepackage[usenames,dvipsnames]{pstricks}
\usepackage{multirow}
\usepackage{tabularx}
\usepackage{cite}
\usepackage{amsmath,amssymb,amsfonts}
\usepackage{graphicx}
\usepackage{textcomp}
\usepackage{xcolor}

\begin{document}
\title{Virtual Node Generation for Node Classification in Sparsely-Labeled Graphs}
%
%
\author{Hang Cui \and
Tarek Abdelzaher}
%
%
\institute{University of Illinois, Urbana Champaign,\\
\email{{hangcui2,zaher}@illinois.edu}}
\maketitle              
\begin{abstract}
  In the broader machine learning literature, data-generation methods demonstrate promising results by generating additional informative training examples via augmenting sparse labels. Such methods are less studied in graphs due to the intricate dependencies among nodes in complex topology structures. This paper presents a novel node generation method that infuses a small set of high-quality synthesized nodes into the graph as additional labeled nodes to optimally expand the propagation of labeled information. By simply infusing additional nodes, the framework is orthogonal to the graph learning and downstream classification techniques, and thus is compatible with most popular graph pre-training (self-supervised learning), semi-supervised learning, and meta-learning methods. The contribution lies in designing the generated node set by solving a novel optimization problem. The optimization places the generated nodes in a manner that: (1) minimizes the classification loss to guarantee training accuracy and (2) maximizes label propagation to low-confidence nodes in the downstream task to ensure high-quality propagation. Theoretically, we show that the above dual optimization maximizes the global confidence of node classification. Our Experiments demonstrate statistically significant performance improvements over 14 baselines on 10 publicly available datasets.
\end{abstract}








\section{Introduction}
In {\em graph literature\/}, the recent state-of-the-art on fine-tuning pre-trained graph models can be summarized into propagation-based methods and infusion-based methods. The propagation-based methods enhance the propagation of graph models via long-range propagation networks~\cite{ding2022meta,liu2023learning} and the discovery of extra propagation patterns~\cite{chen2021topological,tan2022supervised.peng2022self}. Infusion-based methods transfer knowledge from external tasks/domains (often known as base classes) to the new sparse classes (novel classes), such as knowledge graphs~\cite{yao2020graph,dou2022empowering} and meta classes of abundant labels ~\cite{giannone2022scha,ding2022meta,liu2022few,wang2023contrastive}, via meta-learning and semi-supervised methods. Both methods aim to improve the propagation of labeled information to unlabeled nodes.
Previous work suffered from two major drawbacks: (1) Reliance on external knowledge or dataset characteristics: Infusion-based methods require abundant labels from external sources, whereas propagation-based methods rely on dataset characteristics for long-range propagation patterns. (2) Poor adaptation from self-supervised signals. Self-supervised models demonstrate remarkable success on pre-training graph representations prior to downstream tasks. However, previous approaches could not effectively utilize self-supervised signals and are shown to underperform when the labels are sparse ~\cite{tan2022transductive}. 

\begin{figure}
\includegraphics[width=11cm, height=4cm,trim={4cm 10cm 10cm 1cm}, clip]{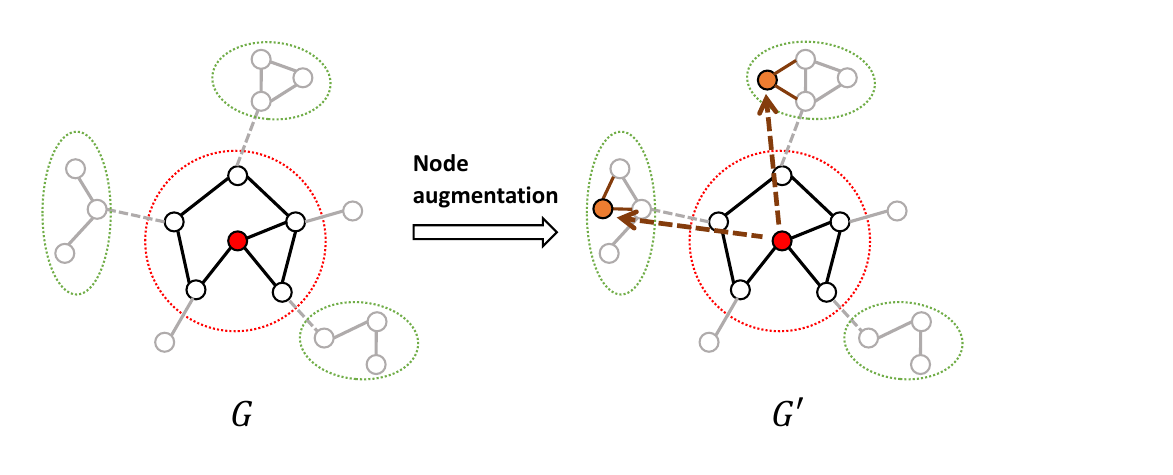}
\caption{Node generation framework: bridging GNN propagation and node augmentation. $G$ represents the original graph where the red circle denotes the sparsely labeled node's (red) local neighborhood, and the green circle denotes low-score regions in the graph. $G'$ is the graph with two generated nodes (orange), which is obtained from augmenting the labeled node (red).}
\label{fig:framework}
\end{figure}

{\em In the broader machine-learning literature\/}, generative methods have shown promising results in numerous machine-learning tasks, including~\cite{duan2021query,liu2022graph,sohn2022robust,yao2021meta,dou2023soft}. The generative methods generate additional informative training examples by augmenting the sparse labels. By operating merely on input data (namely, augmenting the data set), they have the advantage of being orthogonal to and combinable with a variety of other learning methods, including self-supervised, graph meta-learning, and semi-supervised methods. Despite the recent success of generative methods in end-to-end training tasks, there are key challenges in adapting them to graph neural networks (GNN): (1) topology dependence: message-passing models depend on their local neighborhood in the GNN training/inference process; and (2) the over-smoothing dilemma: shallow GNNs have limited propagation ranges, especially when the labels are sparse, while deep GNNs generally suffer from over-smoothing. As a result, the generated nodes must be carefully placed to propagate high-quality information while minimizing over-smoothing. Simple adaptations fail to generate high-quality augmented data in the graph settings without considering the underlying network topology. Therefore, the goal of this paper is to study the \textbf{optimal set of generated nodes that maximizes the propagation of (sparse) labeled information}. 

Our method bridges two popular research areas on graph learning: GNN propagation and node augmentation into an integrated maximum likelihood formulation. Multiple prior efforts~\cite{chen2021topological,tan2022supervised,ding2022meta} have attempted to improve the propagation of labels in a graph, using methods such as edge/substructure inference~\cite{chen2021topological,tan2022supervised} and graph diffusion~\cite{ding2022meta}. However, the availability of such high-quality hyper-relations is often limited. Augmentations~\cite{zhu2020deep,velickovic2019deep,hassani2020contrastive} is frequently used in graph self-supervised learning to improve the robustness of self-supervised representations. This paper bridges the above two pieces of literature by actively exploring high-quality hyper-relations via augmentations, aiming to expand GNN propagation.

Our novel task-aware node generation method is plug-and-play and is compatible with any graph learning method. Given the learned (pre-trained) node representations and downstream node classification task, the goal is to actively generate a small set of high-quality augmented nodes to significantly expand the propagation range of labeled information. We show that for any node-level metrics (i.e., derived from self-supervised signals) for node classification, such as prediction confidence, the set of generated nodes can be optimally determined as a closed-form solution via dual optimizations: (1) Minimize the classification loss of the augmented node; (2) Maximize information propagation to low-score nodes of the node-level metric (i.e., if we use prediction confidence, then the propagation is maximized to low-confidence nodes). The minimization part ensures the classification accuracy of augmented nodes, aiming to reduce cross-class over-smoothing and improve classification accuracy. The maximization part ensures the quality of generated nodes capable of propagating information to key graph regions.

The joint optimization problem can be understood as an optimal `cover' problem on the set of low-score nodes. Although the cover problem is known to be NP-hard, we propose an efficient approximate algorithm that actively searches for the set in a greedy fashion. In addition, we provide several optimizations that further improve our framework in terms of both performance and time complexity.
The contribution of this paper includes:
\begin{itemize}
    \item The first node generation framework to improve node classification in sparsely-labeled graphs, by optimally exploring the set of generated nodes that maximize propagation to low-score areas in the graph.
    \item The method is orthogonal to (and thus combinable with) other popular graph learning techniques, including self-supervised learning and semi-supervised learning, and supplements support and testing sets for graph meta-learning.
\end{itemize}

\section{Related Works}

\subsection{Graph Representation Learning and Node classification}
\textbf{Node classification} is a fundamental problem in graph learning~\cite{cui2018recursive,cui2019semi,cui2021senselens,cui2024unsupervised}. Graph neural networks (GNN)~\cite{zhang2019graph} become the backbone of modern graph learning models. Examples include graph convolutional networks~\cite{zhang2019graph} and graph attention networks~\cite{velivckovic2017graph}. Recently, graph self-supervised learning has emerged as a popular approach, where the graph/node representations are learned unsupervised before knowing the downstream tasks. Popular solutions include graph-autoencoders~\cite{kipf2016variational,hou2022graphmae}, and graph contrastive learning~\cite{zhu2021graph,you2021graph}. Although self-supervised learning provides a promising path towards graph unsupervised learning, it still requires sufficient labeled information to train the classifiers or fine-tuning.

In \textbf{sparsely labeled settings}, previous works attempt to improve the propagation of labeled information to unlabeled nodes, which can be summarized into \textit{propagation-based methods} and \textit{infusion-based methods}. Propagation-based methods explicitly expand propagation by long-range propagation models~\cite{ding2022meta,liu2023learning}, connectivity-based edge sampling~\cite{tan2022supervised}, and topological relational inference~\cite{chen2021topological}. Infusion-based methods transfer knowledge from external sources, such as knowledge graphs and base classes of abundant labels (often known as few-shot meta-learning)~\cite{kim2023task,tan2022transductive,zhou2019meta,wang2022task}. Despite those efforts, previous works cannot effectively utilize self-supervised signals and rely on dataset characteristics for propagation expansion.

\subsection{Data generation for sparse labeled tasks}
Gold-labeled samples are usually limited in real-world tasks due to the high cost of labeling. \textbf{Data-generation-based} methods generate additional training examples by augmenting labeled data. \textit{Feature-oriented methods}~\cite{sohn2022robust} generate diverse but label-invariant synthetic training samples via feature optimization. \textit{Decoupling methods}~\cite{liu2022graph}: decouples task-dependent components and task-invariant components, and then generates training samples via reconstruction. \textit{Interpolation methods}~\cite{yao2021meta} generates training examples via interpolation of labeled data. \textit{Input perturbation}~\cite{ding2022data} performs perturbations on the labeled data, such as masking, rotation, and augmentation.

\section{Motivation and Preliminary}

\subsection{Problem Formulation}

Given an input attribute graph $G(V,E)$, where $V,E$ are the vertex and edge sets, denote the adjacency matrix as $A$ and node feature matrix as $X\in\mathcal{R}^{|V|\times d}$; the goal is to classify unlabeled nodes into $c$ classes, where $c$ is a known parameter. We study the sparse labeled setting, where only a small (e.g., $1\%$) subset of nodes are labeled. Note that, we do not assume any base class with abundant labels. Throughout this paper, we use $X$ as the input features, $Y$ as the labels, and $H$ as the learned embedding. We use $\sim$ to represent node generation variables, for example, $\tilde{V}$ as the set of generated nodes and $\tilde{H}$ as the embedding matrix after node generation.

\subsection{Motivation}
\label{sec:motive}
In machine learning literature, data-generation methods reported state-of-the-art performance in many sparsely labeled machine learning tasks, including named entity recognition~\cite{sohn2022robust}, rationalization~\cite{liu2022graph}, and meta-task generation~\cite{yao2021meta}. The core principle is to generate additional informative training examples from the few labeled samples. The key challenges to applying data generation in graphs lie in the intricate topology dependence of message-passing models and the over-smoothing dilemma of GNNs. Thus the problem is to \textbf{generate an optimal set of synthesized nodes that significantly increases the propagation of labeled nodes to unlabeled nodes}. Our method bridges the two research topics in the following: \\

\noindent\textbf{GNN propagation}\label{sec:propagation}: Previous attempts propose propagating sparse labels information following the principle of homophily: two nodes tend to be similar (share the same label) if connected via a prediction metric $\hat{A} = pred(v_i,v_j)$ that agrees with the label matrix. The prediction metric is often (pre-) trained alongside the GNNs to discover additional propagation patterns. Examples include:
\begin{itemize}
\item Propagation network~\cite{ding2022meta} utilizes diffusion to build the label propagator:
\begin{align}
    \hat{A} = \sum \alpha^{(k)}A^{(k)}, A^{(k)} = TA^{(k-1)}
\end{align}
where $\alpha^{(k)}$ and $T$ are the diffusion parameters.
\item Edge sampling~\cite{tan2022supervised} utilizes link prediction models to generate additional connections:
\begin{align}
    \hat{A} = Connect(V,E)
\end{align}
where $Connect$ is a pre-trained link prediction module such as MLP.
\item Subgraph inference~\cite{chen2021topological} injects new edges among two nodes if the subgraph characteristics of their k-hop neighborhoods are sufficiently close.
\begin{align}
    \hat{A}_{uv} = \frac{exp(d_W(G_u,G_v))}{\sum_{v\in N_k(u)}exp(d_W(G_u,G_v))}
\end{align}
\end{itemize}

The key challenge of the above methods is the limited availability of high-quality connectivity samples in $\hat{A}$.\\

\noindent\textbf{Feature-based node augmentation}: The most commonly used augmentation objective is the classification loss: 
\begin{align}
    \mathcal{L}(f(\tilde{v}),y_{v})
    \label{eq:augment}
\end{align}
where $f()$ is the downstream classifier, $y_v$ is the golden label of node $v$, $\mathcal{L}()$ is a loss function such as cross entropy. The objective measures the classification deviation from augmenting $v$ to $\tilde{v}$. The popular feature-based node augmentation methods~\cite{ding2022data} are:
\begin{itemize}
    \item perturbation-based: $X_{\tilde{v}} = X_v+\delta$
    \item interpolation-based: $X_{\tilde{v}} = (\alpha_1 X_{v_1}+\alpha_2 X_{v_2})/(\alpha_1+\alpha_2)$
    \item mixup-based: mixup features of labeled nodes
\end{itemize}

However, previous augmentation methods in graphs do not explicitly improve the limited propagation range of sparse labels.\\

To counter the above challenges, we propose to bridge the above two lines of research: \textbf{using node augmentations to actively explore high-quality connectivity samples}. The labeled nodes are augmented to satisfy both conditions: (1) align with the downstream classifier in eq.(\ref{eq:augment}); (2) generate high-quality connectivity samples in $\hat{A}$. We show that the augmentations significantly increase the availability of high-quality connectivity samples to expand GNN propagation while also regulated by the augmentation objective to align with the downstream tasks. Our framework is demonstrated in fig.~\ref{fig:framework}.

\section{Our Method}
Given the embedding matrix $H$ from any graph learning methods, denote the downstream classification loss as $\mathcal{L}(f(H),Y)$, where $\mathcal{L}$ is the loss function, $f()$ is the classifier, and $Y$ is the label (ground truth) matrix. Following the discussion in section~\ref{sec:motive}, the labeled nodes are actively augmented to (1) align with the downstream task classifier and (2) generate high-quality connectivity samples:
\begin{align}
   \tilde{V} = argmin_{\tilde{V}} \sum_{\tilde{v}\in\tilde{V}}\mathcal{L}(f(X_{\tilde{v}}),y_v)-\alpha \times Evaluate(\hat{A}|\tilde{V})
   \label{eq:obj}
\end{align}
where the first term is the augmentation objective that minimizes the prediction deviation of augmented nodes $\tilde{v}$ from the original node's label $y_v$; the second term evaluates the propagation quality of $\hat{A}$ generated by the set of augmented nodes $\tilde{V}$. The exact formulation of $Evaluate()$ is derived in the following subsections.
\\

\noindent\textbf{Laplacian smoothness} Recent works~\cite{ma2021unified,liu2023enhancing} demonstrated that the message passing of GNNs can be unified into solving the Laplacian smoothness:

\begin{align}
    H = argmin_H ||H-X||_F^2+\lambda \cdot tr(H^TLH)
\end{align}
where $H$ is the embedding matrix, $X$ is the input feature matrix, $L$ is the normalized Laplacian matrix of the graph, and $\lambda$ is a hyperparameter that controls the smoothness of the graph signal. The above can be rewritten from a node-centric view:
\begin{align}
    H = argmin_H \sum_{i\in\mathcal{V}} ||H_i-X_i||_2^2+\lambda \sum_{i\in\mathcal{V}}\sum_{j\in N_i} ||H_i-H_j||_2^2
\end{align}
where $N_i$ is the set of neighboring nodes of $i$.

The above formulate provides a closed-form derivation of message passing after node generation:
\begin{align}
    \nonumber\tilde{H} = argmin_{\tilde{H}} &\sum_{i\in\mathcal{V}} ||\tilde{H}_i-X_i||_2^2+\lambda \sum_{i\in\mathcal{V}}\sum_{j\in N_i} ||\tilde{H}_i-\tilde{H}_j||_2^2\\
    &+\tilde{\lambda} \sum_{\tilde{i}\in\tilde{V}}\sum_{j\in N_{\tilde{i}}} ||X_{\tilde{i}}-\tilde{H}_j||_2^2
\end{align}
where $\tilde{V}$ denotes the set of generated nodes, $X_{\tilde{i}}$ denotes the features of generated nodes, $\tilde{H}$ represents the feature matrix after node generation ($\tilde{H}$ does not include the generated nodes since they are not part of the classification task).\\

\noindent\textbf{Evaluate propagation of generated nodes} The expected classification performance on unlabeled nodes is commonly estimated as the confidence score $Conf()$ of the classifier~\cite{liu2023semi}. We adopt the increment of confidence score $\Delta Conf()$ (by default reciprocal of variance) as the metric for $Evaluate()$:

\begin{align}
   \nonumber\tilde{V} =& argmin_{\tilde{V}} \sum_{\tilde{v}\in\tilde{V}}\mathcal{L}(f(X_{\tilde{v}}),y_v)\\
   &-\alpha \times [Conf\circ f(\tilde{H})-Conf\circ f(H)]
\end{align}
where $\circ$ denotes function composition.

\begin{proposition}
Given a pre-trained link prediction model $Connect(v_i,v_j)$ and embedding matrix $H$, assume the number of generated nodes is small, s.t. $|V_s|<<|V|$, and the edges of generated nodes are created following Erdos–Renyi  (ER) model proportional to the $Connect(\tilde{v},v_i)$, then the expected $\tilde{H}$ after generating node $\tilde{v}$ with feature $X_{\tilde{v}}$ is:
\begin{align}
    \nonumber\mathbb{E}[\tilde{H}_i] =& \underbrace{(1-\frac{\tilde{\lambda}}{1+d_i\lambda+\tilde{\lambda}}Connect(\tilde{v},v_i))H_i}_\text{original embedding} \\
    &+ \underbrace{\frac{\tilde{\lambda}}{1+d_i\lambda+\tilde{\lambda}}Connect(\tilde{v},v_i)X_{\tilde{v}}}_\text{propagation of $\tilde{v}$}
    \label{eq:H}
\end{align}
where $d_i$ is the degree of node $i$ and $X_{\tilde{v}}$ is the feature of the generated node.
\label{prop1}
\end{proposition}

Proposition.~\ref{prop1} gives a closed-form solution of $\tilde{H}$, which can be understood as an interpolation between the original embedding $H_i$ and the feature of the generated node $X_{\tilde{v}}$.\\

\noindent\textbf{A greedy solution}\begin{proposition}
If the classifier $f()$ and the confidence score metric $conf()$ are convex, and for $\forall \tilde{v}\in\tilde{V}$, $\forall v_i\in V_s$, $Conf\circ f(\tilde{v})>Conf\circ f(v_i)$, then:
\begin{align}
    \sum_{v_i\in V_s}Conf\circ f(\mathbb{E}[\tilde{H}_i])-Conf\circ f(H_i)  
\end{align}
is submodular and monotone in terms of $\tilde{V}$.
\label{prop2}
\end{proposition}

Proposition.~\ref{prop2} grants a greedy approximation of $1-1/\epsilon$, such that the nodes are generated sequentially one at a time:

\begin{align}
   \nonumber\tilde{v} = argmin_{\tilde{v}}& \mathcal{L}_g=\mathcal{L}(f(X_{\tilde{v}}),y_v)\\
   -&\alpha \times \sum_{v_i\in V_s}[Conf\circ f(\mathbb{E}[\tilde{H}_i])-Conf\circ f(H)_i]
   \label{eq:obj}
\end{align}
where $V_s$ is the set of low-confident nodes.

\begin{algorithm}
\caption{A greedy approximation}\label{alg:greedy}
\begin{algorithmic}[1]
\Require Graph $G = (V, E)$, where denotes $V_l$ the set of few-shot labeled nodes, $V_s$ the set of low-confident nodes from downstream classification tasks, $\tilde{V}$ the set of generated synthesized nodes, $\mathcal{L}_g(\cdot)$ the loss function in eq~.(\ref{eq:obj}), $f(\cdot)$ the classifier, $Connect(\cdot,\cdot)$ the pre-trained link prediction model, $Conf(\cdot)$ confident score from downstream classifier, $g(\cdot)$ node augmentation function, hyperparameter for objective loss $\alpha$, learning rate $\tau$, stopping threshold $\delta$, number of generated synthesized node $m$.

\State Initialize $\tilde{V} = \emptyset$
\For{$t=1\cdots m$}
    \For{$v\in V_l$}
    \State $\tilde{v} \gets g_\theta(v)$ 
    \State $grad_{\tilde{v}}(\mathcal{L}_g)\gets\mathcal{L}_g.backward()$
    \While{$grad_{\tilde{v}}(\mathcal{L}_g)>\delta$}
        \State $\tilde{v} \gets \tilde{v} - \tau\cdot grad_{\tilde{v}}(\mathcal{L}_g)$
        \State $grad_{\tilde{v}}(\mathcal{L}_g)\gets\mathcal{L}_g.backward()$
        \EndWhile
    \State $\tilde{V} \gets \tilde{V} + \tilde{v}$ 
    \For{$v_i\in V_s$}
        \State $Conf(v_i)\gets Conf\circ f(\mathbb{E}[\tilde{H}_i])$ 
    \EndFor
    \EndFor
\EndFor
\end{algorithmic}
\end{algorithm}

The detailed pseudocode is shown in Alg.~\ref{alg:greedy}. Line $3-9$ generates an augmented node via gradient descent for a randomly sampled labeled node. The greedy search is initialized with an (optional) pre-defined augmentation function $g_{\theta}(\cdot)$ as discussed in section~\ref{sec:motive}. The optional function enables ad-hoc adaptation from previous data-generation methods for desired characteristics of downstream tasks. After obtaining the set of generated nodes, we generate edges using ER model (proportional to the link prediction probability).\\

\noindent\textbf{Time Complexity} we denote $d_{in}$, $d_h$, and $d_{out}$ as the input, hidden, and output feature size. Assuming $Connect(,)$ is a 2-layer MLP, then the time complexity of our method is $O(|V_s|d^4_{out}d^2_{MLP}+epo\times|E|d_{in}d_{out})$, where $epo$ is the number of epoch of fine-tuning. For larger graphs, the complexity becomes $O(|E|)$.

\subsection{An Simpler Approximation}
Alg.~\ref{alg:greedy} is tesseracted on $d_{out}$, which can be expensive in large-scale practical applications.

By Jensen's inequality,
\begin{align}
    &Conf\circ f(\mathbb{E}[\tilde{H}_i])-Conf\circ f(H_i) \\
    &\leq (1-\frac{\tilde{\lambda}}{1+d_i\lambda+\tilde{\lambda}}Connect(\tilde{v},v_i))Conf\circ f(H_i)\\
    &+\frac{\tilde{\lambda}}{1+d_i\lambda+\tilde{\lambda}}Connect(\tilde{v},v_i)Conf\circ f(\tilde{X}_{\tilde{v}})
\end{align}

The second term can be further relaxed to: 
\begin{align}
\frac{\tilde{\lambda}}{1+d_i\lambda+\tilde{\lambda}}Connect(\tilde{v},v_i)
\end{align}
given that the confidence scores of generated nodes are high. Therefore, eq.(\ref{eq:obj}) is relaxed to 
\begin{align}
&\mathcal{L}_{apx} = \underbrace{\mathcal{L}(f(X_{\tilde{v}}),y_v)}_\text{augment loss}
+\underbrace{\frac{\tilde{\lambda}}{1+d_i\lambda+\tilde{\lambda}}Connect(\tilde{v},v_i)}_\text{edge probability}]\\
&-\alpha \times
\sum_{v_i\in V_s}[\underbrace{(1-\frac{\tilde{\lambda}}{1+d_i\lambda+\tilde{\lambda}}Connect(\tilde{v},v_i))Conf\circ f(H_i)}_\text{propagation to low confident nodes}
\end{align}
$\mathcal{L}_{apx}$ has complexity $O(|V_s|d^2_{out}d^2_{MLP})$. \\

\noindent\textbf{Understand} $\mathcal{L}_{apx}$ has three terms, which govern: (1) augmentation loss of preserving the original class; (2) propagation to low-confident nodes; (3). edge probability between the generated node and the low-confident nodes.

\subsection{Fine-tuning}
After obtaining the generated node-set and the corresponding edges, the graph representation is fine-tuned by predicting the generated node set to its labels. We follow the general data-generation training strategy~\cite{sohn2022robust,liu2022graph, yao2021meta}. Alg.~\ref{alg:greedy} uses multiple reusable components, $Conf(\cdot), g_\theta(\cdot), Connect(\cdot)$, which enables efficient multi-view augmentations. Therefore, we generate multiple sessions of node sets with randomized $V_a$ and initialization, then use a fusion function (such as mean-pooling) to summarize them into the final representation.

\subsection{Optimization and Heuristics}
This section discusses several optimization choices and heuristics:\\
\textbf{Protypical Networks}. The back-propagation of $\mathcal{L}_g$ can be expensive when the set $V_s$ is large. We utilize the prototypical network~\cite{li2020prototypical}, which computes a set of prototypes to summarize $V_s$ through an embedding function, such as k-means. We then only use the prototype set as $\hat{V}_s$ to derive the generated nodes.\\
\textbf{Dynamic pruning}. In most scenarios, we prefer a few highly confident connection samples over many less-confident connections. To achieve this goal, we use a dynamic pruning module that prunes the tail $k_t$ nodes in the link prediction results, every few epochs.\\
\textbf{Hyperparameter}. The hyper-parameter $\alpha$ is vital to balance the trade-off between receptive range and accuracy, which can be tricky to set up. We introduce a dynamical function to update $\alpha$ online. In details, let $W_1 = \mathcal{L}(f(X_{\tilde{v}}),y_v)$ and $W_2 = \sum_{v_j\in V_s}m(\tilde{V},v_j)(1-Conf(v_j))$. We use a controller $Con()$ that stabilizes the ratio of both objectives $W_1/W_2$ to a pre-defined value $\eta$.\\

\section{Experiments}
In this section, we demonstrate our method as a task-aware node generation framework compatible with the existing graph learning methods. For each method, we first obtain their (pre-trained) graph representations, then use our method to generate a set of augmented nodes, and finally fine-tune the graph representation to obtain the classification results.

\subsection{Setup and Pipelines for Self- and Semi-supervised methods}
\textbf{Datasets}
We use 3 small-scale benchmarks~\cite{yang2016revisiting} for node classification: \textit{CiteSeer}, \textit{Cora}, and \textit{PubMed}; and 3 large-scale graphs~\cite{hu2020open}: \textit{ogbn-Arxiv}, \textit{ogbn-Products}, and \textit{ogbn-Product}. For the 3 benchmarks, we use the few-shot setup of $0.5\%$, $1\%$, and $2\%$. For the large-scale graphs, we use $1\%$, $2\%$, and $5\%$. Note that, we do not assume any base classes of abundant labels. In other words, every class in the dataset is considered sparsely labeled. The training and testing split is randomly selected in 10 separate trails. The detailed statistics is shown in table.~\ref{table:dataset}.\\

\begin{table}[]
\small
\centering
\caption{Dataset statistics}
\label{table:dataset}
\begin{tabular}{|l|l l l|}
\hline
Datasets & \#Nodes & \#Edges & \#Features\\\hline
Cora & 2,485 & 5,069 & 1,433\\
Citeseer & 2,110 & 3,668 & 3,703\\
PubMed & 19,717 & 44,324 & 300\\\hline
ogbn-Arxiv & 169,343 & 1,166,243 & 128\\
ogbn-Products & 2,449,029 & 61,859,140 & 100\\
ogbn-Papers100M & 111,059,956  & 1,615,685,872 & 128\\\hline
\end{tabular}
\end{table}

\noindent\textbf{Baselines}
For the three small-scale benchmarks, we compare our methods with self-supervised learning methods: GraphMAE2~\cite{hou2023graphmae2}, GRACE~\cite{Zhu:2020vf}, BGRL~\cite{thakoor2022large}, MVGRL~\cite{hassani2020contrastive}, and CCA-SSG~\cite{zhang2021canonical}; semi-supervised learning methods: GCN~\cite{kipf2016semi}, GAT~\cite{velivckovic2017graph}, Meta-PN~\cite{ding2022meta}, DPT~\cite{liu2023learning}, edge sampling~\cite{tan2022supervised}, and subgraph inference~\cite{chen2021topological}. Unless specified, we run the above baselines with the same experimental setups described in their original papers. 

For the three large-scale datasets, we compare our methods with baselines, which perform well in small-scale benchmarks and include a released implementation scalable to large datasets. The baselines are GraphMAE2, GRACE, BGRL, and CCA-SSG. We also include Simplified Graph Convolution (SGC)~\cite{wu2019simplifying}, a simplified version of GCN for scalability. If available, we use the same source code provided by the above methods; if not, we use GraphSAINT~\cite{zeng2019graphsaint} sampling strategy with GNN initialization. Due to high computation cost, We only apply our node generation to the results of SGC to demonstrate our performance.
\\\\
\textbf{Pipelines}
Since our method does not specify any backbone representation learning method, we use the above baselines to produce the initial representation and then use our method for node generation. To produce a fair comparison, we also fine-tune the self-supervised methods with labeled data on the pre-trained representations. The fine-tuning process uses semi-supervised objectives with the same linear classifiers on default or the objective stated in the baselines if applicable. For semi-supervised models, we use the standard semi-supervised pipelines as default~\cite{sohn2022robust,liu2022graph, yao2021meta}, or the pipelines described in their original papers. 

After the initial classification, we use the reciprocal of prediction variance as the confidence score and generate the set of low-confident nodes $V_s$ using a pre-defined threshold. We use a two-layer MLP as the pre-trained link prediction model. We then output the set of generated nodes (by default 5 per labeled nodes) following eq~.(\ref{eq:obj}) and perform fine-tuning on the representation in 10 separate trials. Note that, the fine-tuning is significantly cheaper as the set of synthesized nodes is small. Finally, we output the classification results using the fine-tuned representation.
\\
\begin{table*}[]
\small
\centering
\caption{Classification Accuracy{\%} of Benchmarks. $+$ presents the fine-tuning results using \textbf{our} proposed method}
\label{table:benchmark}
\hspace*{-0.75cm}\begin{tabular}{|c| c c c | c c c | c c c |} 
 \hline
 Backbone & \multicolumn{3}{c}{Cora} & \multicolumn{3}{c}{CiteSeer} & \multicolumn{3}{c}{PubMed}
 \\  
 \cline{2-10}
 & 0.5\% & 1\% & 2\% & 0.5\% & 1\% & 2\% & 0.5\% & 1\% & 2\%\\
 \hline
 \multicolumn{10}{|c|}{Self-supervised approaches} \\\hline
 GraphMAE2 & $45.2\pm 2.1$ & $55.8\pm 1.4$ & $67.6\pm 1.4$ & $43.7\pm 1.1$ & $56.7\pm 0.9$ & $59.4\pm 0.8$ & $56.0\pm 1.9$ & $62.7\pm 0.8$ & $71.2\pm 2.6$\\
 + & $69.3\pm 1.2$ & $71.8\pm 0.8$ & $78.8\pm 0.6$ & $64.4\pm 0.9$ & $65.5\pm 0.6$ & $66.4\pm 0.6$ & $62.2\pm 1.1$ & $77.1\pm 0.4$ & $78.7\pm 1.0$ \\\hline
 GRACE & $45.8\pm 2.4$ & $62.0\pm 3.6$ & $69.9\pm 1.6$ & $42.6\pm 4.5$ & $52.6\pm 3.6$ & $59.7\pm 3.2$ & $60.9\pm 2.4$ & $67.9\pm 1.4$ & $76.9\pm 3.0$\\
 + & $65.9\pm 1.4$ & $72.7\pm 1.9$ & $79.8\pm 1.1$ & $59.3\pm 1.0$ & $62.3\pm 0.5$ & $67.0\pm 0.4$ & $68.8\pm 1.2$ & $78.8\pm 0.6$ & $83.8\pm 1.2$ \\\hline
 BGRL & $46.6\pm 3.2$ & $60.5\pm 1.4$ & $66.2\pm 2.6$ & $47.9\pm 3.3$ & $52.4\pm 2.7$ & $56.4\pm 2.4$ & $60.4\pm 1.9$ & $66.4\pm 1.1$ & $71.5\pm 2.4$\\
 + & $66.2\pm 1.8$ & $70.6\pm 0.8$ & $74.8\pm 0.4$ & $59.9\pm 1.4$ & $66.0\pm 0.6$ & $71.5\pm 0.4$ & $69.5\pm 1.2$ & $77.6\pm 0.5$ & $81.4\pm 0.4$\\\hline
 MVGRL & $46.4\pm 2.4$ & $60.9\pm 0.9$ & $66.9\pm 2.1$ & $43.2\pm 1.0$ & $53.5\pm 0.8$ & $60.1\pm 1.0$ & $60.2\pm 2.3$ & $68.4\pm 1.5$ & $77.4\pm 2.4$\\
 + & $65.9\pm 1.6$ & $71.4\pm 0.9$ & $75.2\pm 0.8$ & $60.2\pm 0.9$ & $66.5\pm 0.8$ & $71.5\pm 0.5$ & $70.9\pm 1.4$ & $76.9\pm 0.6$ & $80.5\pm 0.4$\\\hline
 CCA-SSG & $48.5\pm 2.0$ & $62.6\pm 1.4$ & $67.9\pm 2.4$ & $47.2\pm 1.4$ & $54.1\pm 1.0$ & $62.2\pm 1.2$ & $61.3\pm 2.0$ & $68.2\pm 0.9$ & $77.5\pm 1.9$\\
 + & $63.2\pm 1.1$ & $70.5\pm 0.6$ & $77.6\pm 0.9$ & $60.3\pm 0.8$ & $61.3\pm 0.6$ & $66.0\pm 0.5$ & $69.9\pm 1.2$ & $77.6\pm 0.5$ & $81.1\pm 0.4$\\\hline
 \multicolumn{10}{|c|}{Semi-supervised approaches} \\\hline
 GCN & $44.2\pm 4.4$ & $48.6\pm 1.8$ & $67.3\pm 2.4$ & $40.2\pm 5.4$ & $53.6\pm 4.0$ & $62.6\pm 2.6$ & $58.5\pm 3.6$ & $66.5\pm 1.5$ & $70.3\pm 4.0$\\
 + & $60.2\pm 2.4$ & $68.1\pm 1.2$ & $73.1\pm 1.9$ & $58.5\pm 2.7$ & $62.8\pm 2.1$ & $69.8\pm 1.4$ & $65.9\pm 2.4$ & $74.2\pm 1.0$ & $79.8\pm 2.6$\\\hline
 GAT & $46.1\pm 4.0$ & $50.4\pm 1.8$ & $69.2\pm 1.4$ & $43.1\pm 5.0$ & $54.2\pm 3.6$ & $63.5\pm 2.4$ & $59.1\pm 3.4$ & $67.0\pm 1.4$ & $70.9\pm 3.4$\\
 + & $61.4\pm 2.0$ & $69.4\pm 1.2$ & $74.9\pm 0.6$ & $59.9\pm 2.2$ & $63.0\pm 1.9$ & $70.0\pm 1.1$ & $66.2\pm 2.0$ & $74.2\pm 0.9$ & $79.6\pm 1.9$\\\hline
  Meta-PN & $50.1\pm 4.0$ & $70.5\pm 1.8$ & $77.6\pm 1.4$ & $50.1\pm 5.0$ & $68.9\pm 3.6$ & $74.2\pm 2.4$ & $71.2\pm 3.4$ & $77.8\pm 1.4$ & $80.5\pm 3.4$\\
 + & $65.4\pm 2.0$ & $74.9\pm 1.2$ & $79.1\pm 0.6$ & $66.2\pm 2.2$ & $70.0\pm 1.9$ & $76.3\pm 1.1$ & $74.1\pm 2.0$ & $79.6\pm 0.9$ & $82.4\pm 1.9$\\\hline
 DPT & $55.2\pm 4.0$ & $70.4\pm 1.8$ & $75.9\pm 1.4$ & $58.1\pm 5.0$ & $68.4\pm 3.6$ & $72.1\pm 2.4$ & $71.4\pm 3.4$ & $77.8\pm 1.4$ & $80.4\pm 3.4$\\
 + & $64.4\pm 1.6$ & $73.1\pm 1.1$ & $78.2\pm 0.6$ & $67.4\pm 1.6$ & $69.8\pm 1.2$ & $75.9\pm 1.1$ & $74.7\pm 1.5$ & $79.1\pm 1.1$ & $81.4\pm 1.6$\\\hline
  Edge & $51.2\pm 5.2$ & $70.5\pm 2.9$ & $74.6\pm 2.6$ & $48.1\pm 5.5$ & $66.4\pm 4.2$ & $73.8\pm 3.8$ & $68.2\pm 4.2$ & $76.2\pm 2.6$ & $80.9\pm 3.2$\\\hline
 Subgraph & $53.6\pm 4.5$ & $71.9\pm 2.6$ & $75.1\pm 2.0$ & $49.1\pm 4.8$ & $67.0\pm 3.8$ & $74.1\pm 3.0$ & $68.3\pm 3.5$ & $76.6\pm 2.5$ & $81.5\pm 2.6$\\\hline

\end{tabular}
\label{table:2}
\end{table*}

\begin{table*}[]
\small
\centering
\caption{Classification Accuracy{\%} of large-scale datasets.}
\label{table:large}
\hspace*{-0.75cm}\begin{tabular}{|c| c c c | c c c | c c c |} 
 \hline
 Baseline & \multicolumn{3}{c}{Arxiv} & \multicolumn{3}{c}{Products} & \multicolumn{3}{c}{Papers100M}
 \\  
 \cline{2-10}
 & 1\% & 2\% & 5\% & 1\% & 2\% & 5\% & 1\% & 2\% & 5\%\\
 \hline
 GraphMAE2 & $66.9\pm 0.5$ & $68.3\pm 0.5$ & $70.2\pm 0.3$ & $77.0\pm 0.4$ & $79.1\pm 0.4$ & $80.5\pm 0.2$ & $58.7\pm 0.4$ & $60.2\pm 0.4$ & $62.9\pm 0.6$\\\hline
 GRACE & $64.3\pm 0.5$ & $66.8\pm 0.4$ & $69.2\pm 0.5$ & $77.1\pm 0.6$ & $78.0\pm 0.6$ & $79.7\pm 0.5$ & $55.5\pm 0.2$ & $57.9\pm 0.2$ & $59.4\pm 0.2$\\\hline
 BGRL & $65.1\pm 1.2$ & $66.6\pm 1.0$ & $69.0\pm 0.3$ & $76.3\pm 0.5$ & $78.0\pm 0.5$ & $79.5\pm 0.4$ & $55.1\pm 0.2$ & $57.9\pm 0.2$ & $60.4\pm 0.5$\\\hline
 CCA-SSG & $64.1\pm 0.2$ & $66.9\pm 0.3$ & $68.3\pm 0.3$ & $75.9\pm 0.4$ & $77.2\pm 0.4$ & $78.5\pm 0.4$ & $55.7\pm 0.2$ & $57.2\pm 0.2$ & $59.8\pm 0.1$\\\hline
 SGC & $63.9\pm 0.5$ & $65.2\pm 0.5$ & $67.0\pm 0.3$ & $73.0\pm 0.4$ & $74.6\pm 0.4$ & $75.6\pm 0.4$ & $59.2\pm 0.2$ & $61.2\pm 0.4$ & $63.9\pm 0.3$\\\hline
 Our & $\textbf{72.1}\pm 0.3$ & $\textbf{73.5}\pm 0.3$ & $\textbf{74.2}\pm 0.3$ & $\textbf{80.6}\pm 0.3$ & $\textbf{81.5}\pm 0.2$ & $\textbf{82.5}\pm 0.2$ & $\textbf{65.5}\pm 0.5$ & $\textbf{66.1}\pm 0.4$ & $\textbf{67.4}\pm 0.4$\\\hline
 
\end{tabular}
\label{table:2}
\end{table*}

\noindent\textbf{Results} The node classification results on the three benchmarks are shown in table.~\ref{table:benchmark}. Our method (represents with `+') improves the node classification performance of self-supervised and semi-supervised learning baselines by a large margin (6.7\%-20.5\%). In addition, our method also reduces the standard deviation of the results over 10 random trials. We observe a larger performance gain on extremely sparsely-labeled scenarios (0.5\%), demonstrating the significance of generating high-quality nodes in a task-aware manner. The fine-tuning results close the gap between the baseline performances, showing our method's independence from any particular pre-training methods. Another notable observation is: self-supervised methods with our fine-tuning techniques can match and outperform semi-supervised methods designed for sparse tasks (Meta-PN and DPT).

Table.~\ref{table:large} shows the node classification results on the three large-scale datasets. Again, we observe large performance gains from our fine-tuning methods, especially under the few-shot scenario(1\%). The overall performances of all the baselines are higher because the datasets are denser. In \textit{Arxiv} and \textit{Papers100M}, we observe a performance gain of (4\%-5.1\%) and (4.5\%-6.8\%), which demonstrates the effective of our method on large datasets. On \textit{Product} however, the performance gain is relatively small (2\%-3.6\%), because the performance is closing to the upper bound of the dataset according to the ogb leaderboard.

\subsection{Computation Time}

\begin{table}[]
\small
\centering
\caption{Running time (s) of training + our fine-tuning. OOM denotes out of memory.}
\label{table:time}
\begin{tabular}{|l|l l l|}
\hline
Baselines & GraphMAE2 & GRACE & GCN\\\hline
Cora & 12.91+0.56(4\%) & 2.76+0.33(12\%) & 0.82+0.11(13\%)\\
Citeseer & 17.89+0.67(4\%) & 4.55+0.42(9\%) & 2.01+0.24(12\%)\\
PubMed & 25.11+1.10(4\%) & 41.3+0.94(2\%) & 4.78+0.49(10\%)\\\hline
Arxiv & 196.8+6.2(3\%) & 612.2+6.1(1\%) & 70.9+0.42(6\%)\\
Products & 1615+20(1\%) & OOM & 311+17(5\%)\\
Papers & 7296+120(2\%) & OOM & 2414+117(5\%)\\\hline
\end{tabular}
\end{table}

This section demonstrates the computation time of pre-training and our node generation fine-tuning. We use baselines: GCN, GraphMAE2 (feature-masking method), and GRACE (graph-augmentation method). The fine-tuning is conducted with 20 epochs for small benchmarks and 50 for large-scale datasets, compared to 200 and 1000 epochs for the initial self-supervised learning. Note that, each epoch of self-supervised learning is also dymatically more expensive than GCN. Table.~\ref{table:time} shows the computation time(s) of GraphMAE2, Grace, and GCN with and without our method on RTX 4090. For larger datasets (ogbn), our approach only adds ~2\% overhead for self-supervised methods and ~5\% for GCN.

\subsection{Ablation Study}

\begin{table*}[]
\small
\centering
\caption{Abalation Study on Components}
\label{table:abalation}
\begin{tabular}{|c|c c c c c c|}
\hline
 & Cora  & Citeseer & PubMed & Arxiv & Products & Papers100M\\\hline
Base & $77.6 \pm 0.6$ & $71.6\pm 0.6$ & $80.9\pm 0.5$ & $74.2 \pm 0.3$ & $82.5\pm 0.2$ & $67.4\pm 0.1$\\\hline
-C.Loss & $75.2 \pm 1.8$ & $69.9\pm 1.6$ & $76.5\pm 2.0$ & $72.8 \pm 1.2$ & $79.9\pm 2.2$ & $65.4\pm 2.0$\\
-Conf() & $74.6 \pm 0.4$ & $67.7\pm 0.6$ & $74.1\pm 0.5$ & $71.9 \pm 0.3$ & $76.4\pm 0.4$ & $62.7\pm 0.4$\\
P.N. & $76.9 \pm 0.5$ & $71.4\pm 0.5$ & $80.8\pm 0.4$ & $73.5 \pm 0.3$ & $80.8\pm 0.4$ & $66.4\pm 0.3$\\\hline
0.01 & $77.5 \pm 0.6$ & $71.0\pm 0.6$ & $81.0\pm 0.6$ & $74.5 \pm 0.3$ & $83.0\pm 0.3$ & $67.0\pm 0.4$\\
0.05 & $77.4 \pm 1.0$ & $71.5\pm 0.6$ & $81.4\pm 0.5$ & $75.2 \pm 0.5$ & $83.1\pm 0.5$ & $68.0\pm 0.9$\\
0.1 & $77.2 \pm 1.2$ & $71.8\pm 1.2$ & $81.0\pm 1.2$ & $76.1 \pm 0.9$ & $82.8\pm 1.2$ & $67.9\pm 1.1$\\\hline
Interpolation & $77.6 \pm 0.6$ & $70.7\pm 0.6$ & $79.2\pm 0.8$ & $75.8 \pm 0.3$ & $80.8\pm 0.2$ & $68.4\pm 0.4$\\
\hline
\end{tabular}
\end{table*}

This section conducts ablation studies to verify the contribution of various components:
\begin{itemize}
    \item Remove classification loss (-C.Loss): the classification loss ensures the generated nodes do not deviate too far from the target class.
    \item Remove $Conf()$ updates at line($12$) (-Conf()). The update is a greedy procedure that reduces the redundancy of the generated node set. 
    \item Use the prototypical network (P.N.) that summarizes the set of less confident nodes.
    \item Using various pre-defined augmentation methods as initialization, using perturbation with parameter $0.01, 0.05, 0.1$; interpolation.
\end{itemize}
We use datasets \textit{Cora}, \textit{Arxiv}, \textit{Papers100M}, and the top-performing model in the previous section as baselines for fine-tuning. The results is shown in table.~\ref{table:abalation}. The classification loss (C.Loss) and greedy procedure (Conf()) both contribute to the performance. The classification loss ensures the generated node stays in safe regions for the classification tasks to reduce classification error and over-smoothing. The greedy procedure reduces the redundant connections to vicinity graph regions. The prototypical network (P.N.) summarizes the less confident areas in the graph to reduce complexity. We observe only minor performance loss for using prototypical networks. Pre-defined augmentation methods show minor performance gains in \textit{Arxiv} and \textit{Papers100M}, which demonstrates the benefit of using prior knowledge of the dataset.

\begin{table}[]
\small
\centering
\caption{Abalation Study on Fine-tuning methods}
\label{table:fine-tuning}
\begin{tabular}{|c|c c c|}
\hline
 & Cora & CiteSeer & PubMed\\\hline
Our method & $73.1 \pm 1.2$ & $69.8\pm 2.1$ & $79.8\pm 1.0$\\\hline
Propagation network & $67.2 \pm 0.9$ & $60.8\pm 1.4$ & $75.4\pm 2.0$\\
edge sampling & $65.5 \pm 1.0$ & $56.0\pm 1.8$ & $73.7\pm 1.2$\\
subgraph inference & $67.1 \pm 1.1$ & $59.2\pm 1.5$ & $75.1\pm 1.3$\\\hline
\end{tabular}
\end{table}

\subsection{Ablation study with Fine-tuning methods}
\label{sec:finetune}
This section compares our method with other few-shot fine-tuning methods described in section.~\ref{sec:propagation}. On default, we use GCN as the baseline as the other methods are built on GCN. The result is shown in table.~\ref{table:fine-tuning}. Our method outperforms the other fine-tuning methods by a large margin (4.4\%-9\%). The other baselines, especially edge sampling, greatly suffer from the limited number of available high-quality connectivities or diffusions. By utilizing node augmentation, our method greatly expands the connectivity's sample space. 

\begin{figure}
\caption{Ablation study on hyperparameters}
\centering
\label{fig:parameters}
\hspace*{-2cm}\includegraphics[width=18cm, height=11cm]{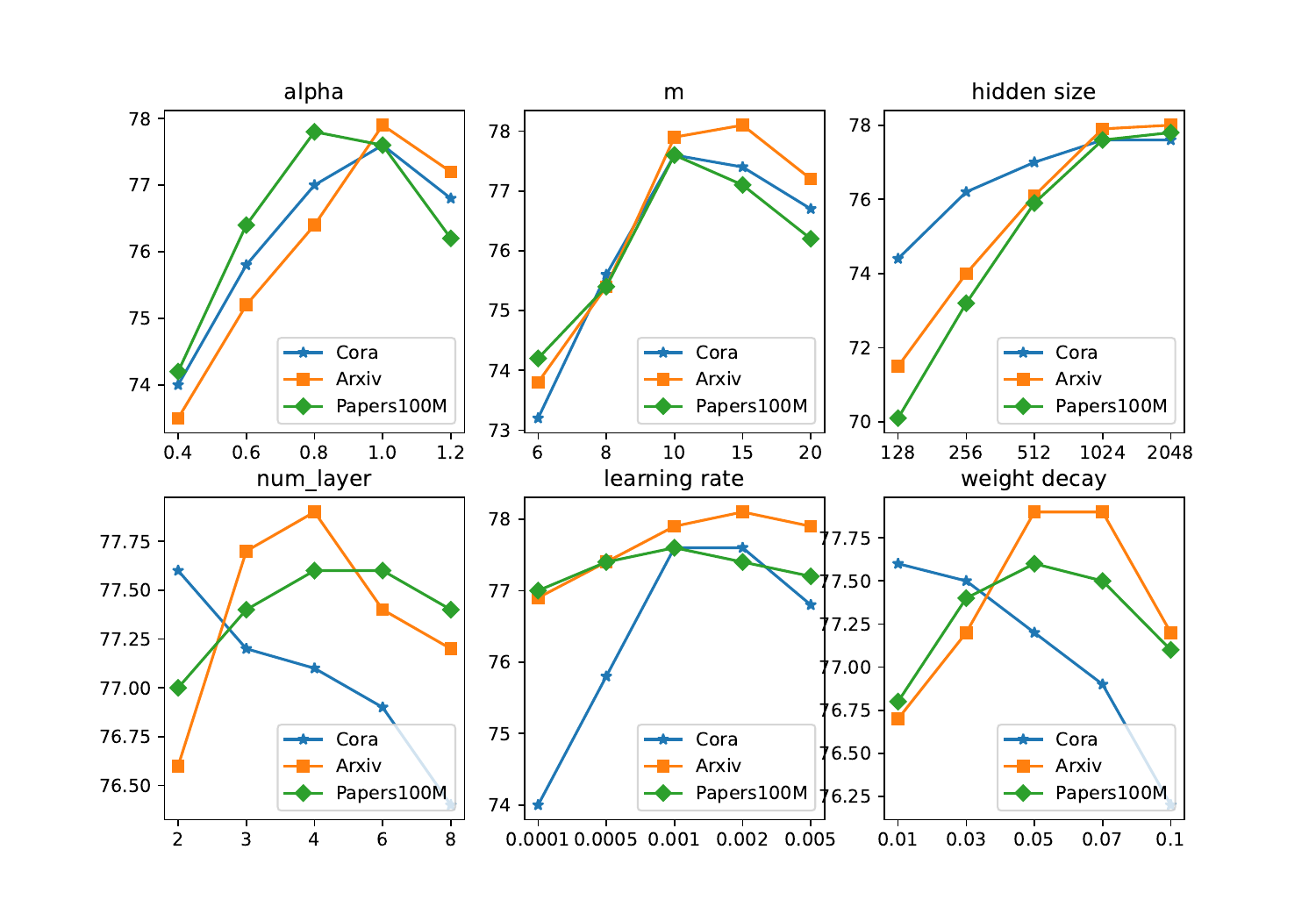}
\end{figure}

\subsection{Parameters}
This section tests our performance on various hyperparameter setups: 
\begin{itemize}
    \item $\alpha$ the trade-off between classification loss and propagation loss in the objective.
    \item $m=|\tilde{V_a}|/|V_a|$ the number of generated nodes per sampled nodes. We constrain the size of generated nodes to reduce over-smoothing. 
    \item model parameters, including hidden size, number of layers, learning rate, and weight decay.
\end{itemize}
Fig.~\ref{fig:parameters} shows the results. In general, our method is not dependent on hyperparameters, with relative performance within $\pm 1.5\%$. The good hyperparameter setup is: $alpha \sim 0.8-1.0$, $m \sim 10-15$, $\text{hidden size}=1024$, $\text{number of layers}\sim 2-4$, $\text{learning rate}\sim 0.001-0.002$, and $\text{weight decay}\sim 0.01-0.05$.

\subsection{Setup and Pipelines for Meta-learning methods}

\begin{table*}[]
\small
\centering
\caption{Classification Accuracy{\%} on 5-way k-shot meta-learning. + presents the results after node generation using our method}
\label{table:meta}
\hspace*{-0.75cm}\begin{tabular}{|c| c c | c c | c c | c c |} 
 \hline
 Backbone & \multicolumn{2}{c}{Corafull} & \multicolumn{2}{c}{Amazon-clothing} & \multicolumn{2}{c}{Amazon-electronics} &\multicolumn{2}{c}{dblp}
 \\  
 \cline{2-9}
 & 3-shot & 5-shot & 3-shot & 5-shot & 3-shot & 5-shot & 3-shot & 5-shot\\
 \hline
 \multicolumn{9}{|c|}{3x labeled nodes in base classes} \\\hline
 TEG & $68.2\pm 2.5$ & $72.4\pm 2.3$ & $82.2\pm 2.5$ & $86.4\pm 2.5$ & $81.6\pm 1.2$ & $86.0\pm 2.3$ & $80.4\pm 1.0$ & $83.9\pm 1.2$\\
 + & $78.5\pm 2.6$ & $80.2\pm 2.7$ & $90.3\pm 2.2$ & $91.2\pm 1.8$ & $87.8\pm 1.8$ & $88.5\pm 1.7$ & $86.9\pm 1.0$ & $86.8\pm 1.2$\\\hline
 TLP & $64.9\pm 2.5$ & $71.9\pm 3.5$ & $83.1\pm 3.5$ & $87.0\pm 3.4$ & $77.6\pm 3.0$ & $83.2\pm 3.2$ & $79.1\pm 3.1$ & $84.8\pm 3.0$\\
 + & $76.0\pm 3.0$ & $79.0\pm 3.0$ & $87.8\pm 2.8$ & $91.5\pm 2.9$ & $88.0\pm 2.6$ & $88.7\pm 2.5$ & $84.6\pm 2.2$ & $86.4\pm 2.2$\\\hline
 Meta-GNN & $55.3\pm 2.4$ & $60.6\pm 3.4$ & $73.7\pm 2.5$ & $76.5\pm 2.1$ & $62.2\pm 1.9$ & $67.8\pm 1.8$ & $68.0\pm 1.9$ & $72.8\pm 2.0$\\
 + & $67.4\pm 2.2$ & $70.4\pm 1.6$ & $83.7\pm 2.2$ & $83.2\pm 1.9$ & $69.7\pm 1.6$ & $70.8\pm 1.6$ & $75.6\pm 1.8$ & $76.9\pm 1.4$\\\hline
  Tent & $62.2\pm 3.1$ & $66.4\pm 2.4$ & $77.1\pm 2.6$ & $82.2\pm 2.6$ & $77.1\pm 1.9$ & $81.1\pm 1.8$ & $79.2\pm 1.9$ & $83.0\pm 1.8$\\
  + & $71.4\pm 2.2$ & $75.5\pm 2.5$ & $86.2\pm 2.6$ & $87.9\pm 1.9$ & $81.0\pm 2.0$ & $84.0\pm 1.9$ & $85.5\pm 1.2$ & $85.8\pm 1.4$\\\hline
 \multicolumn{9}{|c|}{5x labeled nodes in base classes} \\\hline
 TEG & $69.4\pm 2.9$ & $75.5\pm 2.7$ & $87.2\pm 2.2$ & $88.4\pm 2.5$ & $82.2\pm 1.1$ & $86.3\pm 2.3$ & $82.0\pm 1.1$ & $85.9\pm 2.1$\\
 + & $78.3\pm 2.5$ & $80.2\pm 2.4$ & $89.8\pm 1.6$ & $92.7\pm 1.6$ & $90.0\pm 2.0$ & $89.2\pm 1.3$ & $86.2\pm 2.3$ & $87.4\pm 2.6$\\\hline
 TLP & $65.4\pm 3.5$ & $72.4\pm 3.1$ & $83.7\pm 3.4$ & $87.7\pm 3.1$ & $78.1\pm 2.9$ & $85.2\pm 2.8$ & $79.1\pm 2.0$ & $84.4\pm 2.4$\\
 + & $76.4\pm 2.8$ & $79.2\pm 2.8$ & $87.8\pm 2.6$ & $92.7\pm 2.4$ & $88.5\pm 2.6$ & $89.3\pm 2.6$ & $84.5\pm 2.1$ & $86.9\pm 2.4$\\\hline
 Meta-GNN & $57.4\pm 2.6$ & $65.1\pm 3.4$ & $79.7\pm 2.6$ & $80.8\pm 2.2$ & $66.8\pm 2.1$ & $69.1\pm 1.8$ & $71.2\pm 1.9$ & $76.8\pm 1.8$\\
 + & $66.5\pm 2.4$ & $74.4\pm 1.8$ & $88.2\pm 2.2$ & $86.8\pm 1.9$ & $73.8\pm 1.9$ & $72.2\pm 1.8$ & $78.1\pm 1.8$ & $80.8\pm 1.4$\\\hline
  Tent & $63.7\pm 2.1$ & $70.6\pm 2.5$ & $80.2\pm 2.4$ & $84.4\pm 2.5$ & $77.9\pm 1.8$ & $81.5\pm 1.8$ & $81.3\pm 1.9$ & $85.3\pm 2.0$\\
  + & $73.1\pm 2.1$ & $75.4\pm 2.2$ & $86.3\pm 2.2$ & $90.0\pm 1.9$ & $83.4\pm 2.0$ & $84.8\pm 1.6$ & $85.0\pm 1.5$ & $86.4\pm 2.4$\\\hline
  
\end{tabular}
\end{table*}

A key distinction exists between our problem formulation and meta-learning-based methods: we do not assume any base classes with abundant labels, such that the number of labels in every class is few. However, we identify that our node generation approach can be adapted to graph meta-learning by generating additional nodes in training (when labels in base classes are not so abundant) and testing phase. We conduct experiments on four meta-learning baselines: TEG~\cite{kim2023task}, TLP~\cite{tan2022transductive}, Meta-GNN~\cite{zhou2019meta}, and Tent~\cite{wang2022task}.

\noindent\textbf{Datasests} We use popular meta-learning datasets~\cite{kim2023task}: Corafull, Amazon-clothing, Amazon-electronics, and dblp instead of self-supervised learning datasets because meta-learning methods favor datasets with more classes.

\noindent\textbf{Pipelines} As discussed at the beginning of this subsection, meta-learning-based methods require base class with abundant labels while our problem formulation contradicts the above requirement. In particular, if the base class is also k-shot, generating the training episodes is impossible. As a result, we relax our formulation by giving the base classes additional labeled nodes. We conducted three experiments by giving the base classes 3x and 5x labeled nodes. The training episodes are then sampled from the labeled nodes in the base classes. Each method is tested using the released codes and hyperparameters. We use our node generation method to generate additional nodes for base and novel classes (by default 5 per nodes). In detail, we use Grace as the default self-supervised method and then train a logistic classifier on labeled nodes. We then generate additional nodes following alg.\ref{alg:greedy}. We conduct 10 independent trials for each method.

\noindent\textbf{Results} Table.~\ref{table:meta} shows the classification accuracy of meta-learning methods with 3x, 5x, 10x labeled nodes in base classes. Overall, our method improves performance by $\sim 8\%$, $\sim 7\%$, and $\sim 5\%$, respectively. The standard deviation does not reduce as significantly as in previous section due to the random sampling of training episodes. Our experiment stops at 10x labeled nodes because the baseline performance is already similar to abundant labels in base classes.

\section{Conclusion}
This work explores node-generation frameworks for few-shot node classification on graphs. The applications include fine-tuning self-supervised graph representation learning, few-shot semi-supervised learning, and supplementing base and novel classes in graph meta-learning. Our work bridges two lines of research, GNN propagation and graph augmentation, to derive an optimal set of generated nodes that can propagate few-shot information to key areas on the graph. Our framework augments the few labeled nodes to actively explore high-quality connectivity samples to selected low-confident areas on the graph. We propose an efficient greedy solution to approximately find the set and several heuristics to improve the scalability and adaptability of our method. We demonstrate our performance on 10 datasets on 10 baselines.

\section{Proofs}
\subsection{Proposition 1}
\begin{proof}
\begin{align}
H_i &= argmin_{H_i}||H_i-X_i||_2^2+\lambda\sum_{j\in N_i}||H_i-H_j||_2^2\\
&= \frac{1}{1+d_i\lambda}X_i + \sum_{j\in N_i}\frac{\lambda}{1+d_i\lambda}H_j
\end{align}

With probability $\mathbb{P} = 1-Connect(\tilde{v},v_j)$, $\tilde{H}_i = H_i$.\\ 
With $\mathbb{P}=Connect(\tilde{v},v_j)$,

\begin{align}
\tilde{H}_i &= argmin_{H_i}||H_i-X_i||_2^2+\lambda\sum_{j\in N_i}||H_i-H_j||_2^2+\tilde{\lambda}||H_i-X_{\tilde{v}}||_2^2\\
&= \frac{1}{1+d_i\lambda+\tilde{\lambda}}X_i + \sum_{j\in N_i}\frac{\lambda}{1+d_i\lambda+\tilde{\lambda}}H_j+\frac{\tilde{\lambda}}{1+d_i\lambda+\tilde{\lambda}}X_{\tilde{v}}\\
&= \frac{1+d_i\lambda}{1+d_i\lambda+\tilde{\lambda}}H_i+\frac{\tilde{\lambda}}{1+d_i\lambda+\tilde{\lambda}}X_{\tilde{v}}
\end{align}

Thus,
\begin{align}
\mathbb{E}[\tilde{H}_i] &= (1-Connect(\tilde{v},v_j))H_i\\
&+ Connect(\tilde{v},v_j)(\frac{1+d_i\lambda}{1+d_i\lambda+\tilde{\lambda}}H_i+\frac{\tilde{\lambda}}{1+d_i\lambda+\tilde{\lambda}}X_{\tilde{v}})\\
&=(1-\frac{\tilde{\lambda}}{1+d_i\lambda+\tilde{\lambda}}Connect(\tilde{v},v_i))H_i+ \frac{\tilde{\lambda}}{1+d_i\lambda+\tilde{\lambda}}Connect(\tilde{v},v_i)X_{\tilde{v}}
\end{align}

\end{proof}
\subsection{Proposition 2}
\begin{proof}
\textbf{Monotone}. Assume $\tilde{V}$ is ordered in ascending order in terms of $conf\circ f(\tilde{v})$. Let $Conf_i^k$ denote the confident score of node $i$ after generating $k$-th node. Then, we have:
\begin{align}
Conf\circ f(\tilde{v}_{k+1}) \ge Conf_i^{k}
\end{align}
Since the function space of $Conf\circ f$ is convex, we get:
\begin{align}
Conf_i^{k+1} \ge Conf_i^{k}
\end{align}

\textbf{Submodular}. Let $\tilde{V}_s\subset \tilde{V}_t$ and $\tilde{v}\notin\tilde{V}_s, \tilde{v}\notin\tilde{V}_t$. Denote $\tilde{H}^{\tilde{V}_t}$ as the embedding matrix after generating node set $\tilde{V}_t$. $\forall i$, we have

\begin{align}
Conf_i^{\tilde{V}_t} >  Conf_i^{\tilde{V}_s}
\end{align}

\begin{align}
\mathbb{E}[\tilde{H}_i^{\tilde{V}_t+\tilde{v}}]&=(1-\frac{\tilde{\lambda}}{1+d_i\lambda+(|\tilde{V}_t|+1)\tilde{\lambda}}Connect(\tilde{v},v_i))\tilde{H}_i^{\tilde{V}_t}\\
&+ \frac{\tilde{\lambda}}{1+d_i\lambda+(|\tilde{V}_t|+1)\tilde{\lambda}}Connect(\tilde{v},v_i)X_{\tilde{v}}
\end{align}
From the above formulation, we have
\begin{align}
||\mathbb{E}[\tilde{H}^{\tilde{V}_t+\tilde{v}}]-\mathbb{E}[\tilde{H}^{\tilde{V}_t}]|| < ||\mathbb{E}[\tilde{H}^{\tilde{V}_s+\tilde{v}}]-\mathbb{E}[\tilde{H}^{\tilde{V}_s}]||
\end{align}

Together with (29) and convexity:
\begin{align}
    Conf_i^{\tilde{V}_t+\tilde{v}}-Conf_i^{\tilde{V}_t} <  Conf_i^{\tilde{V}_s+\tilde{v}}-Conf_i^{\tilde{V}_s}
\end{align}

\end{proof}

\bibliographystyle{IEEEtran}
\bibliography{main}

\end{document}